\documentclass[12pt]{iopart}
\usepackage{graphicx}
\usepackage{epstopdf}
\usepackage{color}
\usepackage{epsfig}
\usepackage{cite}
\usepackage{amssymb}

\newcommand{\bea}{\begin{eqnarray}}
\newcommand{\eea}{\end{eqnarray}}
\newcommand{\ket}[1]{\left|{#1}\right\rangle}
\newcommand{\bra}[1]{\left\langle{#1}\right|}
\newcommand{\aver}[1]{\left\langle{#1}\right\rangle}

\usepackage{hyperref}

\usepackage{caption} 

\usepackage{enumerate}
\usepackage{float}

\begin{document}
\title{Entanglement dynamics in a model tripartite quantum system}
\author{Pradip Laha, B. Sudarsan, S. Lakshmibala\footnote[1]{Author to whom any correspondence should be addressed.} and \\
V. Balakrishnan}
\address{Department of Physics, Indian Institute of Technology Madras, Chennai 600 036, India}

\begin{abstract}
\paragraph{}
A system comprising a $\Lambda$-type or V-type  atom interacting with two radiation fields  exhibits, during its dynamical evolution, 
interesting  optical phenomena such as electromagnetically-induced transparency (EIT)  and a variety of  nonclassical effects.  Signatures of the latter are seen in the entanglement dynamics of the atomic subsystem and in appropriate  field observables. Some of these effects have been experimentally detected, and have even been used to change the nonlinear optical properties of certain atomic media. It is therefore useful to investigate the roles played by specific initial states of the radiation fields, detuning parameters, field nonlinearities and the nature of field-atom couplings on EIT and on the  entanglement between subsystems. We investigate these aspects in the framework of 
 a simple model that captures the salient features  of such tripartite entangled systems.  Entanglement dynamics is shown to be very sensitive to  
 the intensity-dependent atom-field couplings. Unexpected interesting features pertaining to the collapses and revivals of the atomic subsystem von Neumann entropy appear.  These 
 features   could, in principle, be   useful in enabling   entanglement.
\end{abstract}

 \pacs{ 42.50.-p, 03.67.-a, 03.67Mn, 42.50Dv, 42.50Md}

\section{Introduction}
\label{introduction}

 \paragraph{}
  Interacting quantum systems exhibit many interesting features during temporal evolution. These include diverse  nonclassical effects  such as quantum entanglement,  revival phenomena \cite{robi,aver1,milburn,kita},   collapse of the measure of entanglement 
  to a constant value 
   over certain time intervals,  and so on.   Atom optics provides a convenient framework for examining these effects. In tripartite  entangled systems comprising 
   an atom interacting with two radiation fields, further 
    effects can occur, such as electromagnetically-induced transparency (EIT): the appearance, under suitable conditions, 
     of a transparency window within the absorption  spectrum  of the atomic system. 
      Apart from the change in the transmission coefficient, atomic media can  also exhibit interesting dispersive properties as a consequence of EIT. This 
      feature has been exploited  to create materials that demonstrate slow light-pulse propagation and enhanced nonlinear optical properties (for a review, see \cite{marangos}).
 
 Extensive experimental investigations on EIT have been carried out on three-level atoms interacting with a probe field and a coupling field. Following the report on the occurrence of EIT in optically opaque Sr$^{+}$  vapour in 1991 \cite{boller}, several detailed experiments have been performed on the nature of this optical phenomenon in various atom-field configurations (see, e.g., \cite{Li,entin}).  In the absence of the coupling field, the intensity of the  probe field (equivalently, the corresponding mean photon number)  will remain nearly constant over a small range of values  of the detuning parameter  about zero.  
 The transparency window created by the coupling field in this absorption spectrum is signalled by the appearance of a peak in the probe field intensity.  It is therefore reasonable to expect that, if  the  entangled tripartite system displays collapses and revivals of the probe intensity during time evolution, this peak should be seen at any instant lying in a 
 time interval in which the mean photon number of the probe field collapses to a constant value in the absence of the coupling field.  
  
 In tripartite systems comprising a V-type  or $\Lambda$-type atom and two radiation fields, the inclusion of Kerr-type nonlinearities in the field subsystem 
opens up the possibility of collapses and revivals of the mean photon number corresponding to either field, under specific conditions.  EIT can therefore be investigated, 
for instance,  at an instant  in a time interval when the mean photon number collapses for the first time,  independent of whether or not further  collapses occur.  On longer time scales  the subsystem von Neumann entropy (SVNE) corresponding to the atom can collapse to a constant value over a sufficiently long time interval and this feature could possibly be mirrored in the temporal evolution of the mean photon number of a field subsystem. Further, the manner in which different initial field states, detuning parameters  and interaction strengths affect  both EIT  and the entanglement dynamics on long  time scales needs to be examined. Both  $\Lambda$ and  V
 atoms interacting with a probe field and a coupling field  are good candidates for theoretical and experimental investigations of all  the foregoing  aspects.

The revival phenomenon  is exhibited  even by a single-mode radiation field propagating in  a Kerr-like medium, governed by an effective Hamiltonian  of the form ${a^{\dagger}}^{2} a^{2}$, where $a$ and $a^{\dagger}$ are the photon annihilation 
 and creation operators\cite{milburn, kita}.  Fractional revivals occur at
 specific instants between two successive revivals of the initial wave packet,  when the initial wave packet splits into two or more similar copies of itself \cite{robi}. 
 Signatures  of revivals  and fractional revivals are captured by appropriate quadrature observables  and their higher moments \cite{sudh1, sudh2}.  Collapses and 
 revivals are of greater interest, however, 
  in bipartite and multipartite systems  that involve 
    interaction between the subsystems, entailing nontrivial entanglement dynamics. A bipartite model\cite{puri}  of a multi-level  atom interacting with a single-mode radiation 
    field predicts\cite{sudhjphysb}  collapses and revivals of the  state of the field   (also reflected,  in this case,  in the temporal behaviour of the field SVNE), 
    when the  nonlinearity in the atomic medium  is weak compared to the strength of atom-field interaction.  For stronger  nonlinearity, however, the revival phenomenon is absent;  
    a  detailed time-series analysis of the mean photon  number 
  reveals a gamut  of ergodicity properties  displayed by this observable, 
  ranging from regular to chaotic 
    behaviour\cite{sudhergodicity,sudhrecurrence}. 
 
Entanglement dynamics in the foregoing bipartite model 
turns out to 
 depend  significantly  on the degree to which the initial state of the field
 departs  from perfect coherence.  This has been deduced by selecting, as 
 initial field states, the family of  photon-added coherent states (PACS)\cite{tara} 
 denoted by $\ket{\alpha,m}$, 
 where $\alpha \in \mathbb{C}$ 
 and $m = 0, 1, 2, \ldots\,$.  The standard oscillator coherent state 
 (CS) 
 \begin{equation}
\ket{\alpha}=e^{-|\alpha|^2/2}\,\sum_{n=0}^{\infty}\,
\alpha^{n}\,\ket{n} /\sqrt{n!} 
\label{eqn:cs}
\end{equation}
corresponds to the case $m = 0$, while the 
$m$-photon-added coherent state $\ket{\alpha,m}$
 is obtained by 
 normalising the state $(a^{\dagger})^{m}\ket{\alpha}$ to unity. 
  The set $\{\ket{\alpha,m}\}$ provides a family  of states whose 
departure from coherence is precisely quantifiable.  Experimental realisation 
of the single photon-added coherent state using quantum state 
tomography \cite{zavatta} has  added to the relevance of these studies. 

If the atom has only a very small number of energy levels, 
the dynamics of entanglement  differs both qualitatively and quantitatively from 
that of the bipartite model discussed above.  This aspect has been 
studied\cite{ athreya1, athreya} in the framework of a  
three-level V   or $\Lambda$   atom interacting with a single-mode radiation field. 
A striking feature is that,  relatively independent  of the degree of 
coherence of the initial state of the radiation field, the photon number statistics and the degree  of entanglement are affected strongly  by the low dimensionality of the atomic Hilbert space. This feature continues to hold good for tripartite extensions of the model in which the atom interacts with two radiation fields. 

This  tripartite model is  most suitable   for our  present purposes,  
as it provides a convenient framework for examining EIT, both on  short time scales when the mean photon number records its first collapse,  and on 
longer time scales where  the measure  of entanglement between the atom and fields remains constant  over a significant time interval \cite{athreya1},  for appropriate choices of the field state and parameter values.  Another important aspect  we study 
is the role played by intensity-dependent couplings in entanglement dynamics.  
It has been found \cite{buck} that, for  an  intensity-dependent coupling  of the form 
$(a^{\dagger}a)^{1/2}$  between an initial CS and a Jaynes-Cummings atom, 
the mean photon number can be 
evaluated in closed form, and   the mean photon energy undergoes periodic collapses and revivals in this case.  Subsequent studies have been carried out on 
 the dynamics of the Jaynes-Cummings atom interacting  through an intensity-dependent coupling with other initial states of the radiation field such as the squeezed vacuum  and the $SU(1,1)$ coherent state \cite{buzek1, buzek2}.  A  coupling proportional to  $1/(a^{\dagger}a)^{1/2}$ has 
 also been used\cite{zait} 
 in a tripartite model,  motivated by the fact that this form arises 
 naturally in the context of diagonal-state representations of the 
 density matrix in a restricted Hilbert space where the zero-photon state is 
 absent\cite{sudarshan}.  An intensity-dependent coupling 
 of the form  $(1 + \kappa\,a^{\dagger}\,a)^{1/2}, \; 0 \leq \kappa \leq 1$ 
 has  been shown\cite{siva1}  to lead to  a closed-form expression for the mean photon energy.   Here  $\kappa = 0$ reduces to the Heisenberg-Weyl algebra 
 for  the field operators,  while $\kappa = 1$ leads to the  
 $SU(1,1)$ algebra for (nonlinear combinations 
 of) these operators\cite{siva2}. Intermediate values of 
 $\kappa$  corresponds to a deformed $SU(1,1)$ operator algebra. 
 While the revival phenomenon has been examined in several  models with 
 intensity-dependent couplings, including the tripartite system of an atom interacting with two radiation fields  \cite{tav1, tav2}, the effect  of such a coupling on the 
entanglement  dynamics and on EIT  has not been investigated. In this paper, 
we report on the effects of a general  intensity-dependent coupling of the form  
$(1 + \kappa \:{a^{\dagger}}a)^{1/2}$ on these two phenomena  in the 
tripartite model  of a 
$\Lambda$  atom  interacting with two radiation modes. We have also investigated the corresponding case of a V  atom. We do not present those  results, as 
they  are not qualitatively 
different in any significant manner from those for the $\Lambda$  atom.

 \section{$\Lambda$  atom interacting with two radiation modes}
 \label{model}
 
\paragraph{} 
 
 The tripartite model has two radiation fields: a 
 `probe field'  $F_{1}$ and a `coupling field' $F_{2}$, of 
 respective frequencies 
 $\Omega_{1}$ and $\Omega_{2}$, with annihilation and creation 
 operators $a_{i}$ and  $a^{\dagger}_{i}\,(i = 1,2)$. 
   The highest energy  state of the  $\Lambda$ atom is denoted by  $\ket{3}$, 
    and $\ket{1}$ and $\ket{2}$ are the two lower energy states. $F_{1}$ 
    and $F_{2}$ induce, respectively,  the $\ket{1}\leftrightarrow \ket{3}$ 
    and $\ket{2}\leftrightarrow \ket{3}$ transitions, while the  transition $\ket{1}\leftrightarrow \ket{2}$ is 
    dipole-forbidden.  The general Hamiltonian that incorporates field nonlinearities and intensity-dependent couplings is given (setting $\hbar = 1)$ by 
 \begin{eqnarray}
\fl {H}  =  \sum\limits_{j=1}^{3} \omega_{j} \sigma_{jj}  +  \Omega_{1} \,a_{1}^{\dagger} a_{1}  +  \chi_{1} \,a_{1}^{\dagger 2} a_{1}^{2}  +  \Omega_{2} \,a_{2}^{\dagger} a_{2}  +  \chi_{2} \,a_{2}^{\dagger 2} a_{2}^{2} \nonumber \\
 +  \lambda_{1} ( R_{1} \,\sigma_{31} +  R^{\dagger}_{1} \,\sigma_{13} )  +  \lambda_{2} ( R_{2} \,\sigma_{32}  +  R^{\dagger}_{2}\,\sigma_{23} ).
\label{eqn:two_mode_lambda_hamiltonian}
\end{eqnarray}
Here, $\sigma_{ij} = \ket{i}\bra{j}$ where $\ket{j}$ is an atomic state, $\{\omega_{j}\}$ are positive constants, $\chi_{i}$ represents the strength of the  
nonlinearity in the field $F_{i}$,  and $\lambda_{1}, \lambda_{2}$ are the 
respective atom-field coupling strengths  corresponding to the $\ket{3}\leftrightarrow \ket{1}$ and $\ket{3}\leftrightarrow \ket{2}$ transitions.   Further, 
\begin{equation}
\fl R_{i}  =  a_{i} \,f(N_{i}), 
\label{defnofR}
\end{equation}  
 where $f(N_{i})$ is  a real-valued function of $N_{i} = a^{\dagger}_{i} a_{i}$ 
 that serves to incorporate a possible  intensity-dependent coupling. 
As stated  in the Introduction, we consider  the functional 
 form $f(N_{i} )= (1 + \kappa_{i} N_{i})^{1/2}$, 
where  $\kappa_{i}$ takes values  in the range  $[0,1]$.  Wherever relevant, 
a comparison will be made with results pertinent to  the  cases 
 $f(N_{i}) = 1$ and $f(N_{i}) = N_{i}^{1/2}$.  
$H$ can be written as the sum $H_{0} + H_{1}$, where  
\numparts
\begin{eqnarray}
\fl {H}_{0}  =  \sum\limits_{i=1}^{2}  \Omega_{i} N_{i}^{\rm tot}  +  \omega_{3} I,
\label{eqn:two_mode_lambda_h0}
\end{eqnarray}
\begin{eqnarray}
\fl {H}_{1}  =   \sum\limits_{i=1}^{2} \chi_{i} a_{i}^{\dagger 2} a_{i}^{2} - \Delta_{i} \sigma_{ii} + \lambda_{i} ( R_{i} \sigma_{3i}  +  R^{\dagger}_{i} \sigma_{i3} ).
\label{eqn:two_mode_lambda_h1}
\end{eqnarray}
\endnumparts
Here $I  =  \sum_{j=1}^{3} \sigma_{jj}$, $N_{i}^{\rm tot}  =  a_{i}^{\dagger} a_{i}  -  \sigma_{ii} \;(i=1,2)$  are constants of the motion,  and  the two detuning parameters $\Delta_{i}$ are given by
\begin{equation}
\Delta_{i} = \omega_{3} - \omega_{i} - \Omega_{i}.
\label{eqn:two_mode_v_delta1}
\end{equation}
$H_{0}$ merely introduces a phase factor in the time evolution of the state. 
It is therefore convenient to work in 
an appropriate interaction picture to eliminate this trivial dependence.

Earlier work \cite{tav2} pertaining to  a $\Lambda$ 
 atom interacting with two radiation fields 
 involved an intensity-dependent coupling specified by $f(N_{i}) =  
 N_{i}^{1/2}$, and  a  cross-Kerr nonlinearity between the fields, but not individual field nonlinearities as in the present instance.  Entanglement collapse to a constant value over a  significant time interval has not been reported hitherto, nor has EIT been examined in this case.  The model we consider  here provides a natural setting for examining the effects of  individual field nonlinearities, detuning parameters, a family  of 
 intensity-dependent couplings and specific classes of  initial states on 
 EIT and  entanglement  dynamics   during temporal evolution.

 The quantities we require for our study are the matrix elements of the 
 time-dependent 
 reduced density matrices of the tripartite system. 
  We consider $F_{1}$ and $F_{2}$  to be initially in general 
 superpositions of  Fock states, $\sum_{0}^{\infty} q_{n} \ket{n}$ 
 and $\sum_{0}^{\infty} r_{m} \ket{m}$, respectively. For definiteness,  the initial state of the atom is taken 
 to be  $\ket{1}$ throughout. The initial state of the full system is thus 
\begin{equation}
\fl \ket{\psi (0)}  =  \sum\limits_{n=0}^{\infty} \sum\limits_{m=0}^{\infty} q_{n} r_{m} 
\ket{1; n; m}, 
\label{eqn:two_mode_lambda_initial_state}
\end{equation}  
where $\ket{1; n;  m}$ denotes a state with the 
 atom in $\ket{1}$ and the fields 
 $F_{1}$ and $F_{2}$ in  states with $n$ and 
 $m$ photons, respectively. (An analogous notation will be used for 
 other field states as well, such as coherent and photon-added 
 coherent states.)  Proceeding  on  lines similar 
to those for the V  atom\cite{athreya},  the interaction picture 
state  vector at time $t$  is  found to be 
\begin{eqnarray}
\fl \ket{\psi (t)}_{\rm int} = \sum\limits_{n=0}^{\infty}\sum\limits_{m=0}^{\infty} q_{n} r_{m} \biggl\{ A_{nm}(t) e^{i \Delta_{1}t} \ket{1; n; m} \nonumber \\
 + \,B_{nm}(t) e^{i \Delta_{2}t} \ket{2; n-1; m+1}  \nonumber \\
  + \,C_{nm}(t) \ket{3; n-1; m} \biggr\},
\label{eqn:two_mode_lambda_interaction_state}
\end{eqnarray}
where the 
functions $A_{nm}(t), B_{nm}(t), C_{nm}(t)$ are as follows.  
Since the atom is taken to be initially in 
the state $\ket{1}$, it cannot make a transition to $\ket{3}$ 
if $n = 0$. Hence  
\begin{equation}
\fl 
A_{0m}(t)  = 1, \;B_{0m}(t)  = 
C_{0m}(t)  = 0
\label{eqn:two_mode_v_a_n0}
\end{equation}
for all $m$.  When  $n, m \geq 1$,  we find  
\numparts
\begin{eqnarray}
\fl A_{nm}(t)  =  \frac{e^{i(\Delta_{2} - \Delta_{1}) t}} {f_{1}f_{2}} \sum\limits_{j=1}^{3} 
b_{j} \bigg\{ (\Delta_{2} + \mu_{j}  +  V_{12}  +  V_{22})  \nonumber \\
+ (\mu_{j}  +  V_{12}  +  V_{21})  -  f_{2}^{2} \bigg\}  e^{i \mu_{j} t},
\label{eqn:two_mode_lambda_a}
\end{eqnarray}
\begin{equation}
\fl B_{nm}(t)  =  \sum\limits_{j=1}^{3} b_{j} e^{i \mu_{j} t},
\label{eqn:two_mode_lambda_b}
\end{equation}
\begin{equation}
\fl C_{nm}(t)  = -  \frac{e^{i \Delta_{2} t}} {f_{2}} \sum\limits_{j=1}^{3} b_{j} (\mu_{j}  +  V_{12}  +  V_{21}) e^{i \mu_{j} t},
\label{eqn:two_mode_lambda_c}
\end{equation}
\endnumparts
where 
\numparts
\begin{equation}
\fl V_{11}  =  \chi_{1} n (n-1), \quad V_{12}  =  \chi_{1} (n-1)(n-2),
\label{eqn:two_mode_lambda_v11v12}
\end{equation}
\begin{equation}
\fl V_{21}  =  \chi_{2} m(m+1), \quad V_{22}  =  \chi_{2} m (m-1),
\label{eqn:two_mode_lambda_v21v22}
\end{equation}
\begin{equation}
\fl f_{1}  =  \lambda_{1} n^{1/2} f(n), \quad f_{2}  =  \lambda_{2} 
(m+1)^{1/2} f(m+1).
\label{eqn:two_mode_lambda_f1f2}
\end{equation}
\endnumparts
The effect of the operator $f(N)$ in the Hamiltonian  on the state 
of the system
is captured in the quantities $f_{1}$ and $f_{2}$.  Further, 
\begin{equation}
\fl \mu_{j}  =  - {\textstyle\frac{1}{3}}\,x_{1}  +  
{\textstyle\frac{2}{3}} \,(x_{1}^{2} -  3 x_{2})^{1/2} 
\cos \,\Big\{\theta  +  
{\textstyle\frac{2}{3}} (j-1) \pi  \Big\}
\label{eqn:mu_j}
\end{equation}
where $j  =  1, 2, 3$, and 
\begin{equation}
\fl \theta  =  {\textstyle\frac{1}{3}}  
\cos^{-1} \Big\{ [9 x_{1} x_{2} -  2 x_{1}^{3}  -  27 x_{3}]
\big/[2 ( x_{1}^{2}  -  3 x_{2} )^{3/2}]
\Big\},
\label{eqn:theta}
\end{equation}
\numparts
\begin{eqnarray}
\fl x_{1}  =  V_{11}  +  2 V_{12}  +  V_{21}  +  2 V_{22} - \Delta_{1} + 2\Delta_{2},
\label{eqn:two_mode_lambda_x1}
\end{eqnarray}
\begin{eqnarray}
\fl x_{2} =  (V_{12}  +  V_{21} + \Delta_{2}) (V_{11}  +  V_{12}  +  2 V_{22} - \Delta_{1}) \nonumber \\
+  (V_{12}  +  V_{22}) (V_{11}  +  V_{22} - \Delta_{1})  + 2\Delta_{2}(V_{12} + V_{21})  \nonumber \\
 + \Delta_{2}^{2} - f_{1}^{2}  -  f_{2}^{2},
\label{eqn:two_mode_lambda_x2}
\end{eqnarray}
\begin{eqnarray}
\fl x_{3}  = \Delta_{2}(V_{12} + V_{21})(V_{11} + V_{12} + 2V_{22} - \Delta_{1}) \nonumber \\
 -  f_{2}^{2} (V_{11}  + V_{22} - \Delta_{1} + \Delta_{2})+ \Delta_{2}^{2}(V_{12} + V_{21})  \nonumber \\
 + (V_{12}  +  V_{21}) \left\{ (V_{12}  +  V_{22}) (V_{11}  +  V_{22} - \Delta_{1})  -  f_{1}^{2} \right\}, 
\label{eqn:two_mode_lambda_x3}
\end{eqnarray}
\endnumparts
and
\begin{equation}
\fl  b_{j}  = f_{1} f_{2}/[(\mu_{j}-\mu_{k}) (\mu_{j}-\mu_{l})], \;
j \neq k \neq l.
\label{eqn:b_j}
\end{equation}  
Finally, for  $m=0$ we find 
\numparts
\begin{equation}
\fl A_{n0}(t) = \sum\limits_{j=1}^{2} c_{j} e^{i \alpha_{j} t},
\label{eqn:two_mode_v_a_m0}
\end{equation}
\begin{equation}
\fl B_{n0}(t) = 0,
\label{eqn:two_mode_v_b_m0}
\end{equation}
\begin{equation}
\fl C_{n0}(t) = - \frac{e^{i \Delta_{1}}} {f_{1}} \sum\limits_{j=1}^{2} c_{j} (\alpha_{j}  +  V_{11}) e^{i \alpha_{j} t},
\label{eqn:two_mode_v_c_m0}
\end{equation}
\endnumparts
where 
\begin{equation}
\fl 
 c_{1}  =  \frac{V_{11} + \alpha_{2}} {\alpha_{2}  -  \alpha_{1}},\quad c_{2}  =  \frac{V_{11} + \alpha_{1}} {\alpha_{1}  -  \alpha_{2}}
\label{eqn:two_mode_v_a1}
\end{equation}
and 
\begin{equation}
\fl  \alpha_{1,2} = {\textstyle\frac{1}{2}}
 [-y_{1} \pm (y_{1}^{2} - 4y_{2})^{1/2}],
\label{eqn:two_mode_v_a1}
\end{equation}
 where 
\begin{equation}
\fl y_{1}  =  V_{11}  +  V_{12},\quad y_{2}  =  V_{12} V_{11}  -  f_{1}^{2}.
\label{eqn:two_mode_v_y1}
\end{equation}
In the expressions for $B_{nm}(t)$ in the foregoing, the contribution 
from spontaneous emission has not been included, as it can be 
shown to be negligible.

With these results at hand,  
the   general matrix elements of the reduced density 
matrices $\rho_{1}(t), \rho_{2}(t)$ and $\rho_{\rm a}(t)$  
for the  field subsystems $F_{1}, F_{2}$ and 
the atom subsystem,  respectively, 
can be computed by tracing over the other two subsystems. 
We obtain, finally, 
\begin{eqnarray}
\fl \bra{n} \rho_{1} (t) \ket{n^{\prime}} = \sum_{l=0}^{\infty} \bigg[ q_{n} q_{n^{\prime}}^{*} r_{l} r_{l}^{*} A_{n,l} A_{n^{\prime},l}^{*} + +  q_{n+1} q_{n^{\prime}+1}^{*} r_{l} r_{l}^{*} C_{n+1,l} C_{n^{\prime}+1,l}^{*}\nonumber \\ 
+ (1-\delta_{l,0})\big( q_{n+1} q_{n^{\prime}+1}^{*} r_{l-1} r_{l-1}^{*} B_{n+1,l-1} B_{n^{\prime}+1,l-1}^{*} ) \bigg],
\label{eqn:two_mode_lambda_rho_f1_matrix_elts}
\end{eqnarray}
\begin{eqnarray}
\fl \bra{l} \rho_{2} (t) \ket{l^{\prime}} = \sum\limits_{n=0}^{\infty} \bigg[ q_{n} q_{n}^{*} r_{l} r_{l^{\prime}}^{*} A_{n,l} A_{n,l^{\prime}}^{*} +  q_{n+1} q_{n+1}^{*} r_{l} r_{l^{\prime}}^{*} C_{n+1,l} C_{n+1,l^{\prime}}^{*} \nonumber \\
 + (1-\delta_{l,0}) (1-\delta_{l^{\prime},0})\big( q_{n+1} q_{n+1}^{*} r_{l-1} r_{l^{\prime}-1}^{*} B_{n+1,l-1} B_{n+1,l^{\prime}-1}^{*} ) \bigg],
\label{eqn:two_mode_lambda_rho_f2_matrix_elts}
\end{eqnarray}
and 
\numparts
\begin{eqnarray}\label{eqn:two_mode_lambda_rho_at_matrix_elts_a}
\fl \bra{1} \rho_{\rm a} (t) \ket{1} &=& \sum\limits_{k=0}^{\infty} \sum\limits_{l=0}^{\infty} q_{k} q_{k}^{*} r_{l} r_{l}^{*} A_{k,l} A_{k,l}^{*}, \\ 
\label{eqn:two_mode_lambda_rho_at_matrix_elts_b}
\fl \bra{2} \rho_{\rm a} (t) \ket{2} &=& \sum\limits_{k=0}^{\infty} \sum\limits_{l=1}^{\infty} q_{k+1} q_{k^+1}^{*} r_{l-1} r_{l-1}^{*} B_{k+1,l-1} B_{k+1,l-1}^{*}, \\
\label{eqn:two_mode_lambda_rho_at_matrix_elts_c}
\fl \bra{3} \rho_{\rm a} (t) \ket{3} &=& \sum\limits_{k=0}^{\infty} \sum\limits_{l=0}^{\infty} q_{k+1} q_{k+1}^{*} r_{l} r_{l}^{*} C_{k+1,l} C_{k+1,l}^{*}, \\
\label{eqn:two_mode_lambda_rho_at_matrix_elts_d}
\fl \bra{1} \rho_{\rm a} (t) \ket{2} &=& \sum\limits_{k=0}^{\infty} \sum\limits_{l=1}^{\infty} q_{k} q_{k+1}^{*} r_{l} r_{l-1}^{*} A_{k,l} B_{k+1,l-1}^{*}, \\ 
\label{eqn:two_mode_lambda_rho_at_matrix_elts_e}
\fl \bra{1} \rho_{\rm a} (t) \ket{3} &=& \sum\limits_{k=0}^{\infty} \sum\limits_{l=0}^{\infty} q_{k} q_{k+1}^{*} r_{l} r_{l}^{*} A_{k,l} C_{k+1,l}^{*}, \\ 
\label{eqn:two_mode_lambda_rho_at_matrix_elts_f}
\fl \bra{2} \rho_{\rm a} (t) \ket{3} &=& \sum\limits_{k=0}^{\infty} \sum\limits_{l=1}^{\infty} q_{k+1} q_{k+1}^{*} r_{l-1} r_{l}^{*} B_{k+1,l-1} C_{k+1,l}^{*}.
\end{eqnarray}
\endnumparts
The quantity of interest in the context of EIT (to be analysed in 
the next section) is the 
expectation value  of the  photon number of the  probe field $F_{1}$, given by
\begin{equation}
\fl  \aver{N_{1}(t)} = \Tr\,[\rho_{1}(t) N_{1}],   
\label{}
\end{equation}
while entanglement dynamics (examined in section \ref{entanglement}) 
 is characterised by the SVNE of the atomic 
subsystem, defined as
\begin{equation}
\fl S_{\rm a}(t)  = - \Tr\, [\rho_{\rm a}(t) \,\ln\,\rho_{\rm a}(t)].
\label{}
\end{equation}

  \section{Electromagnetically-induced transparency (EIT)}
  \label{eit}
  
 \paragraph{}  
 We are ready, now, to investigate  EIT for a  $\Lambda$  
  atom interacting with the probe field $F_{1}$ and the coupling field $F_{2}$.   
In order to have a reference point from which to 
understand the general behaviour, we consider first 
 the relatively simple  case in which  the detuning  
 parameter $\Delta_{2}=0$, the couplings $\lambda_{1} = \lambda_{2} = \lambda = 1$, and the field nonlinearity parameters $\chi_{1} = \chi_{2} = 0$.  As mentioned earlier, we consider situations  where  $\aver{N_{1}}$ exhibits collapses and revivals  during 
 the temporal evolution of the system,  and  examine 
 features of  EIT during that particular interval of time when $\aver{N_{1}}$  first collapses to a constant value in the absence of  $F_2$.

\begin{figure}[h]
\centering
\includegraphics[scale=0.85]{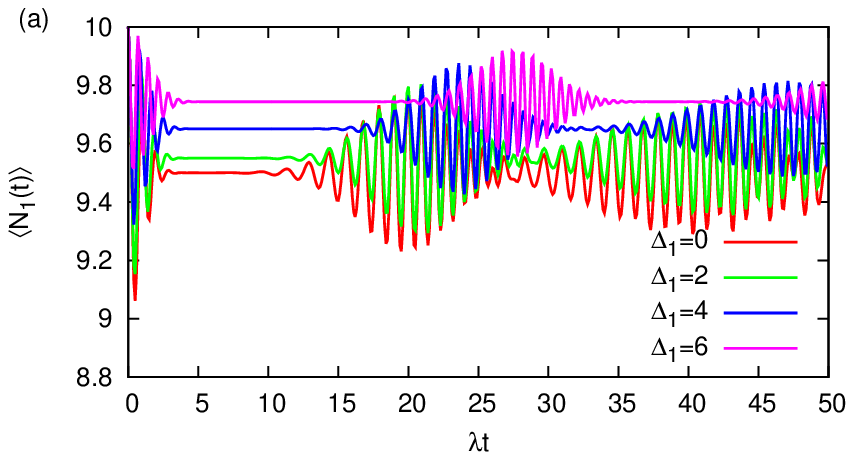}
\includegraphics[scale=0.85]{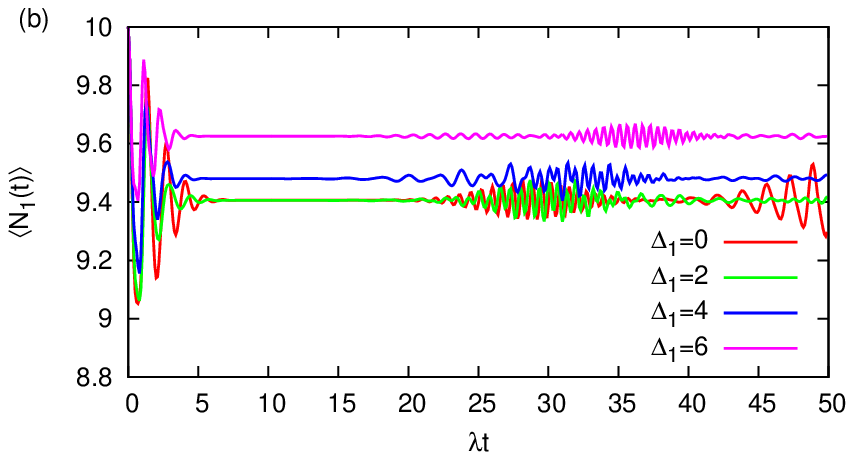}
 \includegraphics[scale=0.85]{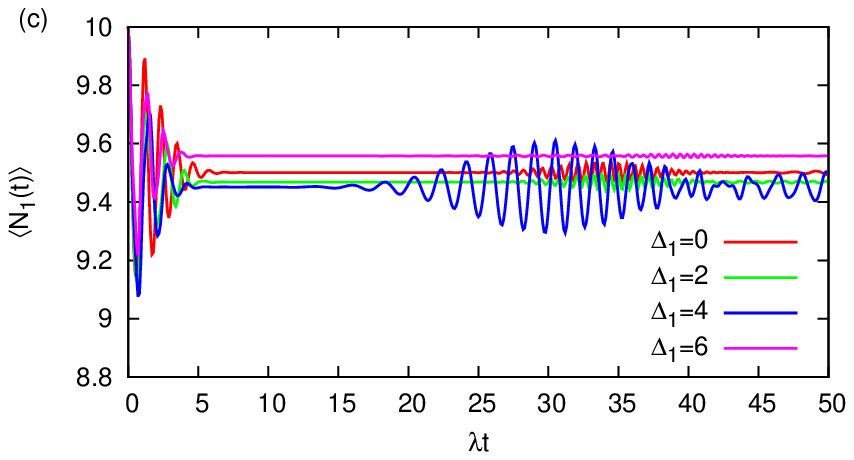}
\includegraphics[scale=0.85]{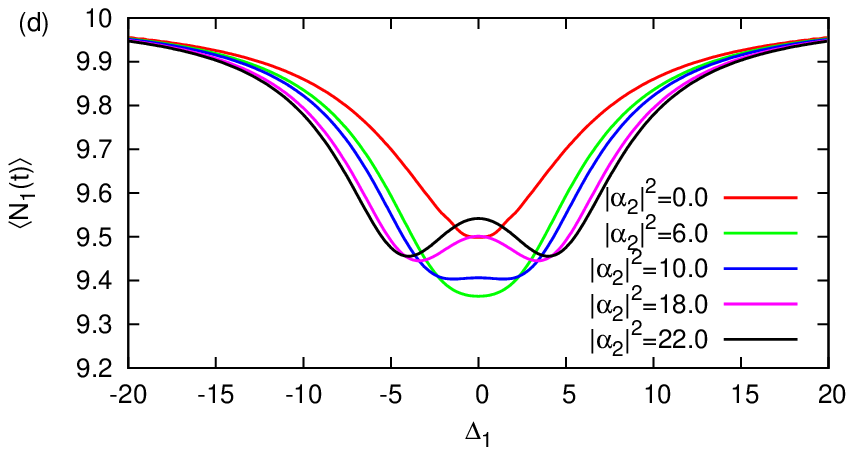}
\caption{$\aver{N_{1}(t)}$ in the case of  initial  coherent  states  
$\ket{\alpha_{1}}$ and $\ket{\alpha_{2}}$ of the field modes. 
$|\alpha_1|^2=10$ and (a) $|\alpha_2|^2=0$, (b) $|\alpha_2|^2=10$,  (c) 
$|\alpha_2|^2=18$. (d) $\aver{N_{1}}$ as a function of  the detuning parameter 
 $\Delta_{1}$, for different values of $|\alpha_{2}|^{2}$.}
\label{fig:EIT_1}
\end{figure}
When the initial states of 
$F_1$ and $F_2$ are coherent states 
$\ket{\alpha_{1}} $ and $\ket{\alpha_{2}}$, 
$\aver{N_{1}(t)}$ undergoes collapses and revivals in time for a  range of values of  $\Delta_1$ and $|\alpha_2|^{2}$, for 
a given value of $|\alpha_1|^{2}$ (figures  \ref{fig:EIT_1}(a)-(c)). The time interval over which $\aver{N_{1}}$ exhibits collapse is sensitive to the values of  both $\Delta_{1}$ and $|\alpha_1|^{2}$.  Since the intervals of first collapse  for different values of  $\Delta_{1}$ overlap with each other, we can capture the appearance of EIT  in a plot of $\aver{N_{1}(t)}$ versus $\Delta_{1}$ at any specific instant of time in this overlap interval (figure \ref{fig:EIT_1}(d)). 
In particular, the generation of a transparency window around $\Delta_1=0$ for $|\alpha_{2}|^2 \gg |\alpha_{1}|^2$ is manifest. 
We have also verified that the occupation probabilities 
of the atomic states $|1\rangle$ and $|3\rangle$  
appropriately complement  the behaviour of $\aver{N_{1}}$  in this time interval.
\begin{figure}[h]
\centering
\includegraphics[scale=0.85]{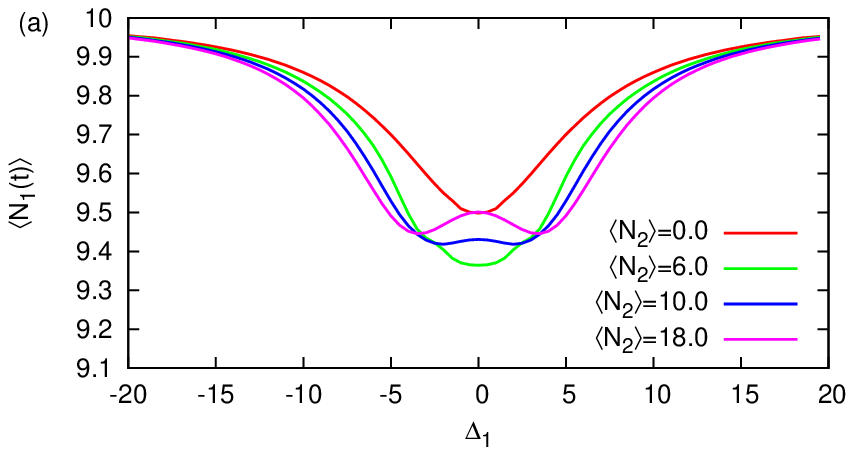}
\includegraphics[scale=0.85]{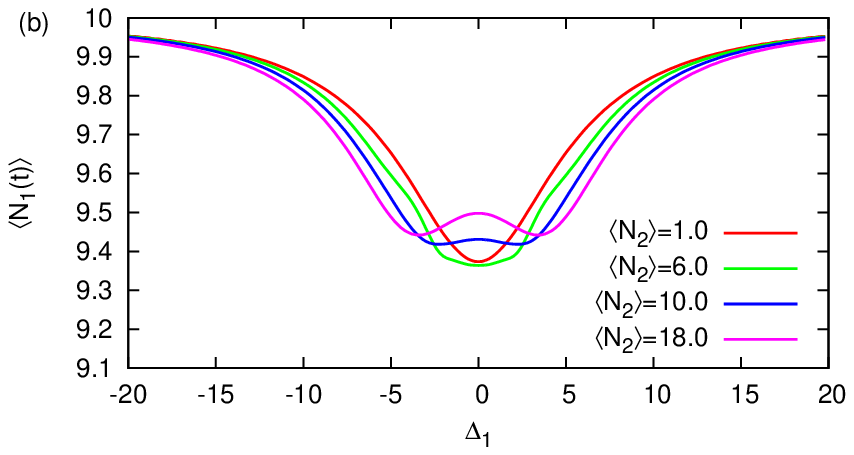}
\includegraphics[scale=0.85]{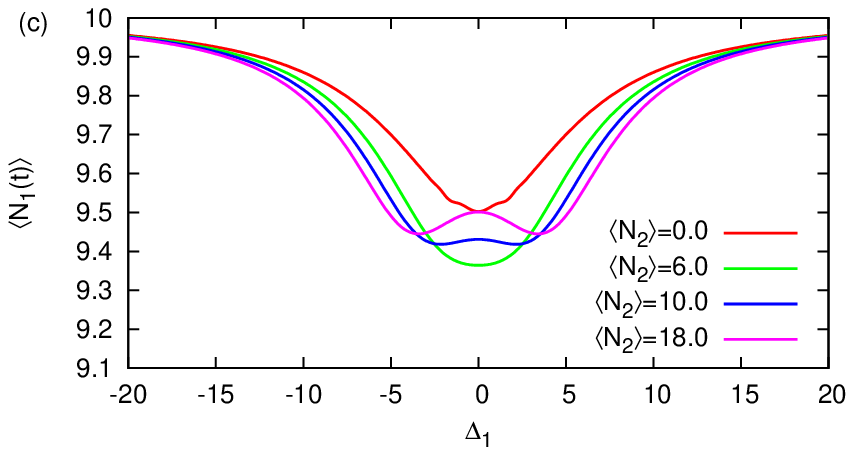}
\includegraphics[scale=0.85]{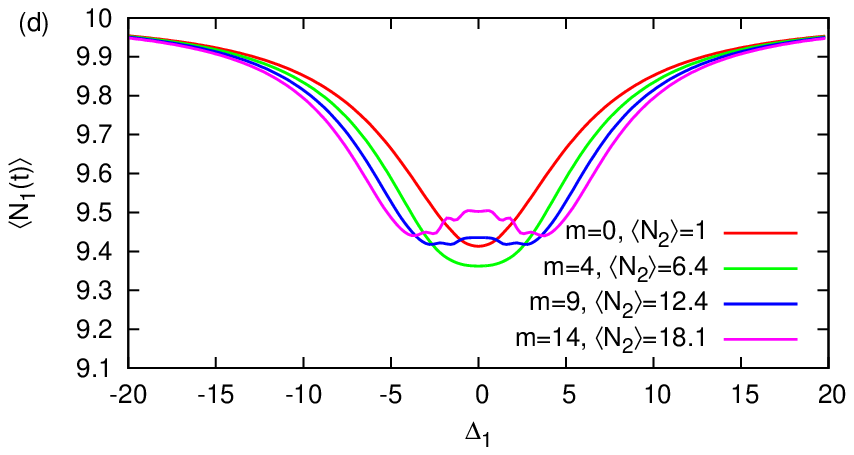}
\caption{$\aver{N_{1}(t)}$ versus $\Delta_{1}$, when the  initial state of $F_{2}$ is   
(a) an even CS (b) an odd CS, (c) a Yurke-Stoler state  (d) an $m$-photon-added 
CS.} 
\label{fig:EIT_catpacs}
\end{figure}
 
 We now proceed to examine the effect of field nonlinearities, intensity-dependent couplings and  different initial states of the coupling field on EIT.  
 As before,  we take $F_1$ to be 
 initially in a CS 
 with $|\alpha_{1}|^2=10$, and  set $\chi_1 = \chi_2 = 0$, $\Delta_2=0$, $\lambda_1=\lambda_2=\lambda=1$, while 
 the initial state of $F_2$ is  chosen  to be an 
 $m$-PACS or a Schr\"{o}dinger cat state.  Once again,  
 EIT  occurs during the first interval of collapse of $\aver{N_{1}(t)}$  (figures  \ref{fig:EIT_catpacs}(a)-(c)). The height of the transparency window in the absorption spectrum  now depends directly on $\aver{N_{2}}$, consistent with the experimental finding \cite{boller}. 
 When the initial state of  $F_2$ is  an $m$-PACS (as opposed to a CS), then,    
 with increasing  $m$,  the field mode undergoes 
 incomplete collapses which have 
 Rabi-like oscillations about a mean value. This is reflected in the jaggedness of the transparency  peak for large values of $m$ (figure \ref{fig:EIT_catpacs}(d)). For very large values of 
 $m$ the  intervals of collapse corresponding to  different values of $\Delta_1$ 
 no longer overlap significantly,  consistent with  what happens when
    $F_2$ is  initially in a photon number state.

\begin{figure}[h]
\centering
\includegraphics[scale=0.85]{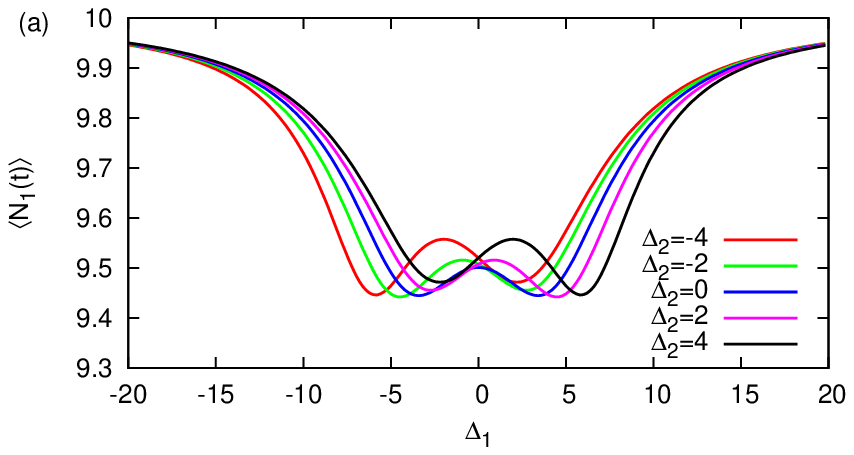}
\includegraphics[scale=0.85]{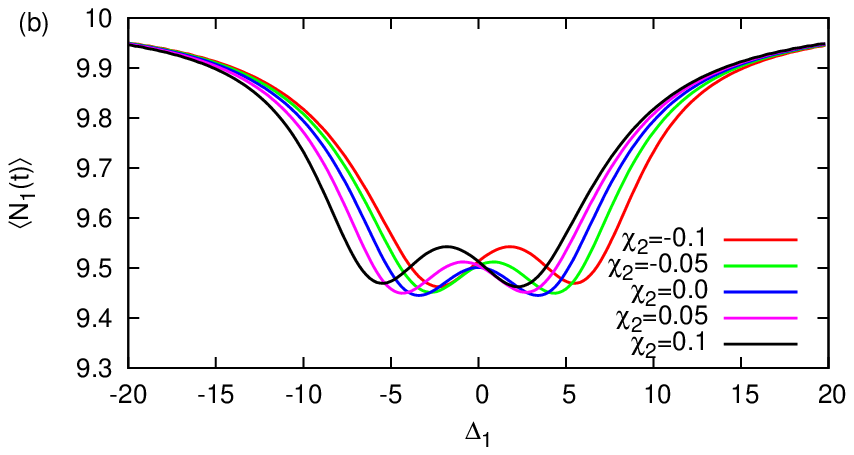}
\caption{EIT spectrum for different values of (a) $\Delta_2$ and (b) $\chi_2$.}
\label{fig:EIT_kerrsweep2}
\end{figure}

  The presence of nonlinearities in $F_1$ mitigates collapses of $\aver{N_{1}(t)}$. On the other hand, if  $F_2$  has a Kerr-like nonlinearity, 
   complete collapses in $\aver{N_{1}(t)}$ occur.  We  
   investigate the nature of EIT during the 
   time interval of the first collapse by choosing 
 initial  coherent states with $|\alpha_{1}|^2 = 10$ and  
   $|\alpha_{2}|^2 = 18$ for $F_{1}$ and $F_{2}$, 
     setting $\lambda_1=\lambda_2=\lambda=1$, $\chi_1=0$,   and varying  $\chi_2$. 
  The EIT spectrum (a plot of $\aver{N_{1}(t)}$ versus $\Delta_1$)  
     is no longer   symmetric about $\Delta_{1}=0$,  in contrast to the earlier case, 
     and the asymmetry 
      becomes more pronounced  with increasing  $\chi_2$. 
      A noteworthy feature is that this behaviour is similar to that 
      obtained by varying $\Delta_2$ over a small range of 
      values about zero (figures 
  \ref{fig:EIT_kerrsweep2}(a), (b)).

 \begin{figure}
\centering
\includegraphics[scale=0.85]{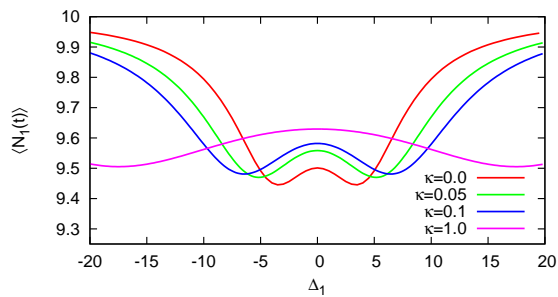}
\caption{EIT spectrum for different values of $\kappa$.} 
\label{fig:EIT_idc}
\end{figure}

  Finally, if the coupling between the atom and either of the two fields has  
  an intensity dependence of the form $f(N) = 
 (1+\kappa N)^{1/2}$,  
 the EIT spectrum stretches and flattens out  
 with increasing $\kappa$.  Figure \ref{fig:EIT_idc} illustrates this 
 feature in the case of  initial coherent  states of $F_{1}$ and $F_{2}$ with  
 $|\alpha_{1}|^2=10, |\alpha_{2}|^2=18$. Unlike what happens for 
 non-zero values of  $\chi_{2}$ or $\Delta_{2}$, however, the 
 spectrum remains symmetric about 
 $\Delta_{1} = 0$ as $\kappa$ is increased,  
 while the minima in $\aver{N_{1}(t)}$ that  indicate EIT  
 become less pronounced and  move out to larger and larger 
 values of $|\Delta_{1}|$.

\section{Entanglement dynamics of a $\Lambda$  atom coupled to 
two  field modes}
\label{entanglement}

\paragraph{} 
 
  We turn, next, to the entanglement dynamics  of our tripartite  system. 
  We start  with the simple case of an intensity-independent  field-atom coupling, and 
  then go on to  consider the case of an
   intensity-dependent coupling proportional to 
    $(1+\kappa N)^{1/2}$.  In order to avoid inessential complications, 
     we  set $\Delta_{1}=\Delta_{2} =\Delta$, $\lambda_{1}=\lambda_{2}=\lambda$, $\chi_{1}=\chi_{2}=\chi$, and $\kappa_{1}=\kappa_{2}=\kappa$.
 
\subsection{Field-atom interactions with constant coupling strengths }
\begin{figure}[h]
\centering
\includegraphics[scale=0.85]{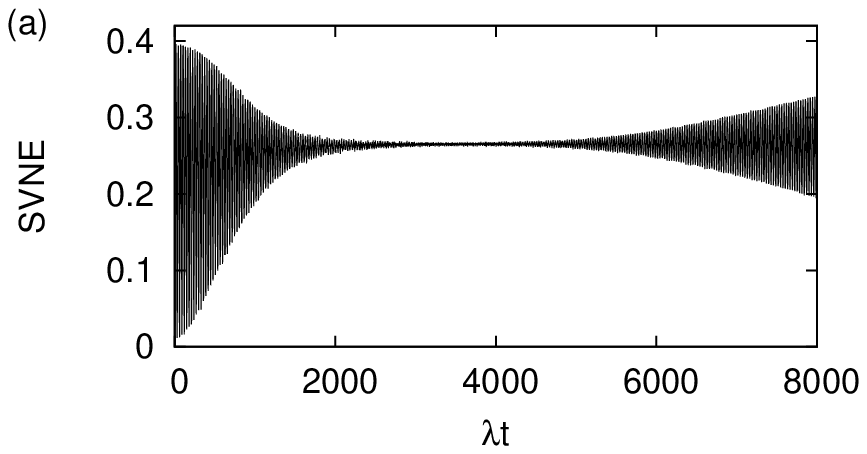}
\includegraphics[scale=0.85]{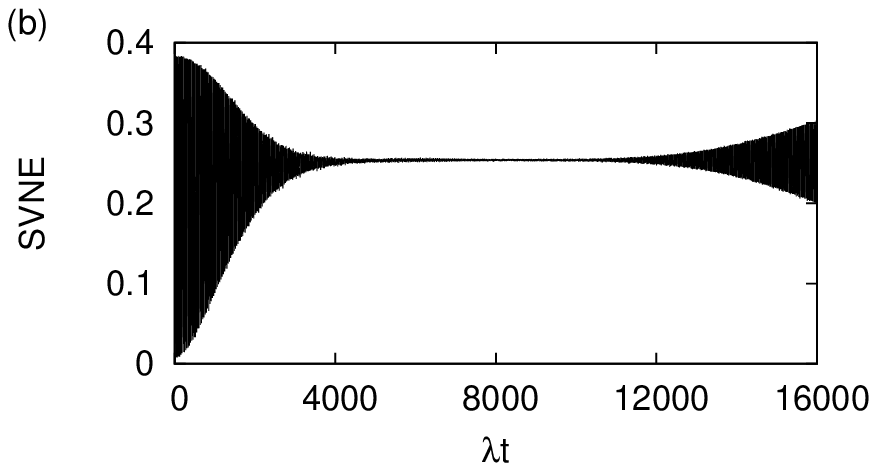}
\caption{$S_{\rm a}(t)$  for a 
$\Lambda$ atom interacting with two field modes, 
for  $\chi/\lambda = 5$,   
$\Delta = 0$.   
Initial state (a) $\ket{1;\alpha, 5; \alpha, 5}$ and  
(b) $\ket{1;\alpha, 10; \alpha, 10}$, with 
${|\alpha|}^{2} =10$.} 
\label{fig:svne_1}
\end{figure}

\paragraph{} 
Earlier  work\cite{athreya1} has indicated an interesting feature in the 
behaviour of the  entanglement 
in the system as measured by   $S_{\rm a}(t)$, the SVNE
of  the atomic subsystem, 
when the nonlinearity 
parameter $\chi$ is large compared to the field-atom interaction strength $\lambda$. 
In this regime, $S_{\rm a}$  displays a collapse (to a steady value) 
when  the initial field 
state is a PACS, in contrast to what happens when it is a CS. 
This feature has been verified in detail for the system at hand, and 
figures \ref{fig:svne_1}(a) and (b) depict some representative results 
in this regard.   With an increase in the 
number of added photons in the initial field states, there is a 
systematic  increase  of the 
interval during which  $S_{\rm a}$ remains at a steady value.  
 We have verified that these features are also reflected in the Mandel Q parameter
 and in the  mean and variance of the photon number  $N_{1}$.

\begin{figure}[h]
\centering
\includegraphics[scale=0.85]{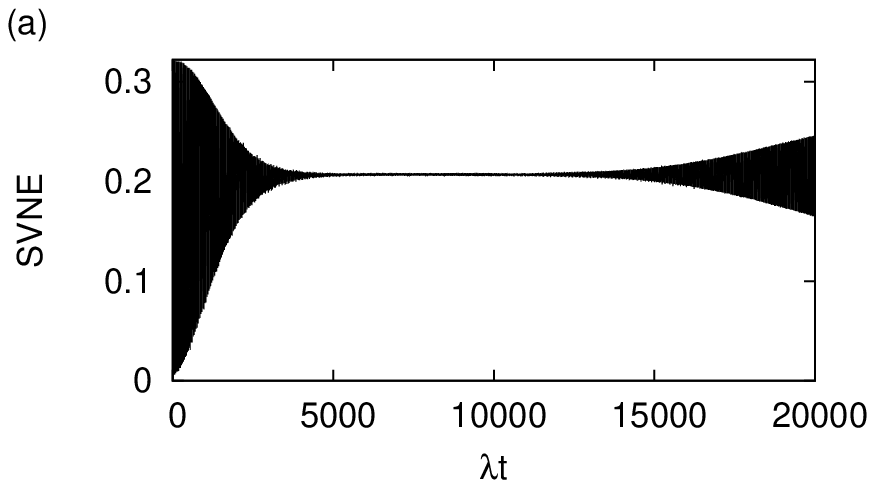}
\includegraphics[scale=0.85]{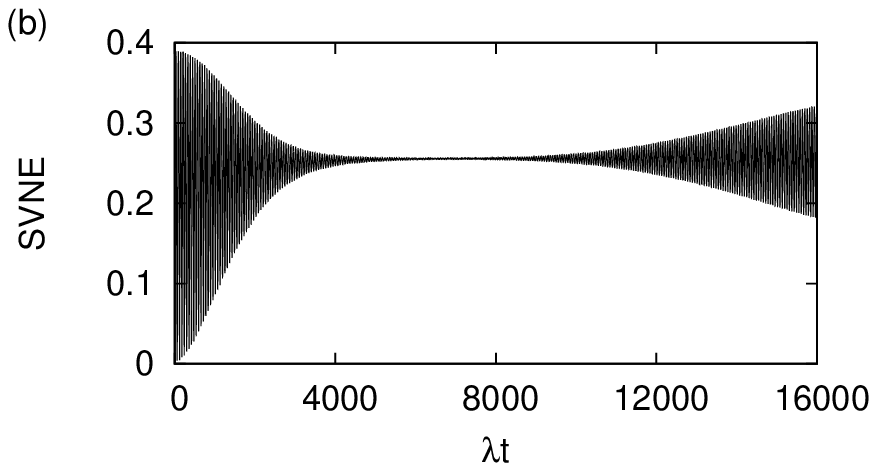}
\caption{$S_{\rm a}(t)$ for 
the initial state $\ket{1;\alpha, 5; \alpha, 5}$,  where (a) ${|\alpha|}^{2} = 20$ 
and $\chi/\lambda = 5$;    (b) ${|\alpha|}^{2} =10$ and $\chi/\lambda = 10$ 
(strong nonlinearity). 
In both cases, $\Delta = 0$.} 
\label{fig:svne_2}
\end{figure}

\begin{figure}[h]
\centering
\includegraphics[scale=0.85]{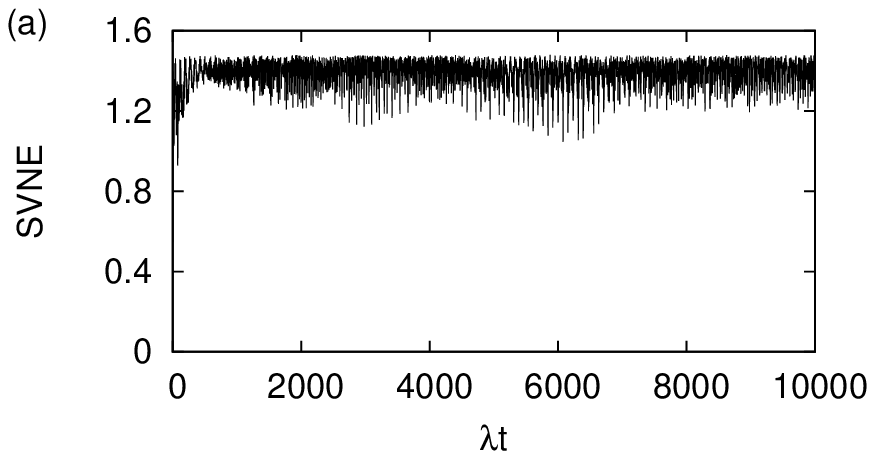}
\includegraphics[scale=0.85]{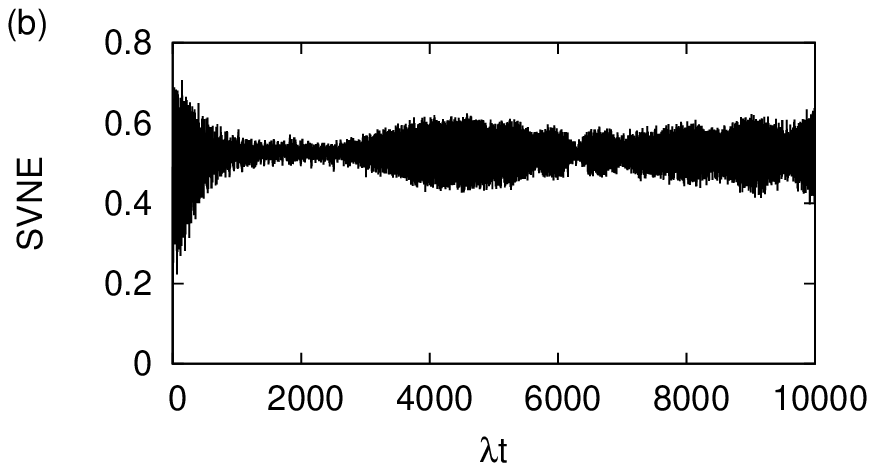}
\caption{$S_{\rm a}(t)$ 
for the  initial state $\ket{1;\alpha, 5; \alpha, 5}$ with 
 ${|\alpha|}^{2} = 10$ and  (a) $\chi = 0$ and  (b) $\chi/\lambda = 1$ (weak nonlinearity). 
 In both cases, $\Delta = 0$.} 
\label{fig:svne_3}
\end{figure}

The  time interval during which the SVNE 
holds at a steady value is enhanced  when 
 ${|\alpha |}^{2}$ is increased 
 (compare figures \ref{fig:svne_1}(a) and \ref{fig:svne_2}(a)), 
 or    the ratio ${\chi}/{\lambda}$ is increased
(see, e.g.,  figures \ref{fig:svne_1}(a) and \ref{fig:svne_2}(b)), or both, 
 all other parameters remaining unchanged.  
This sort of entanglement collapse is absent in  the case of  
initial coherent states,  and also for initial PACS provided the  
nonlinearity parameter $\chi$ is sufficiently 
small \cite{athreya1}. The route to entanglement collapse with increase in nonlinearity is evident by comparing figures \ref{fig:svne_3}(a) and (b)  
with figure \ref{fig:svne_1}(a).  As one might expect,  there is a 
shrinkage of the time interval over which the entanglement collapses  
as one moves away from  exact resonance to non-zero values of the 
detuning parameters $\Delta_{i}$. 
 
\begin{figure}[h]
\centering
\includegraphics[scale=0.57]{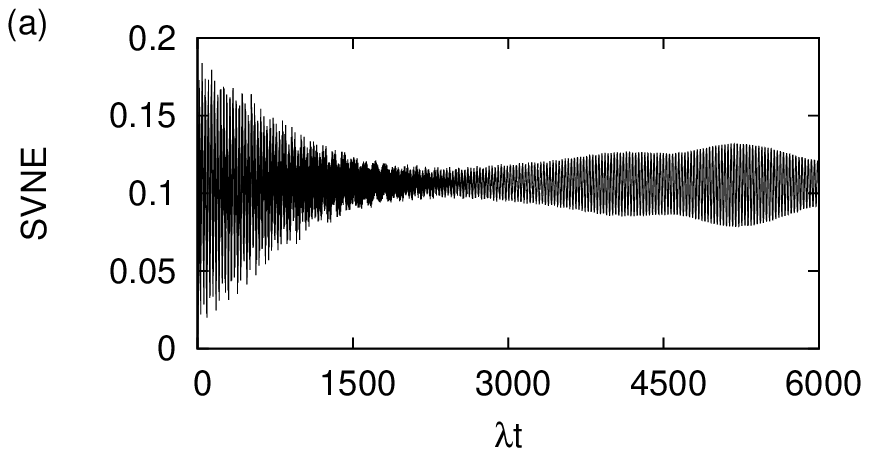}
\includegraphics[scale=0.57]{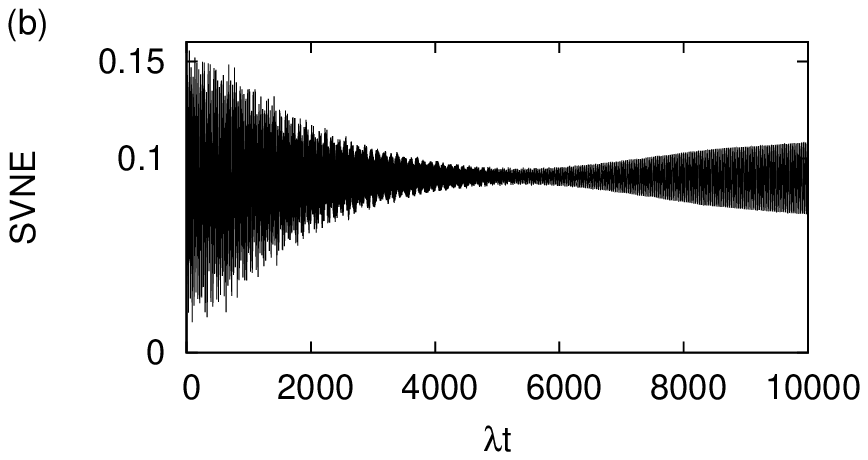}
\includegraphics[scale=0.57]{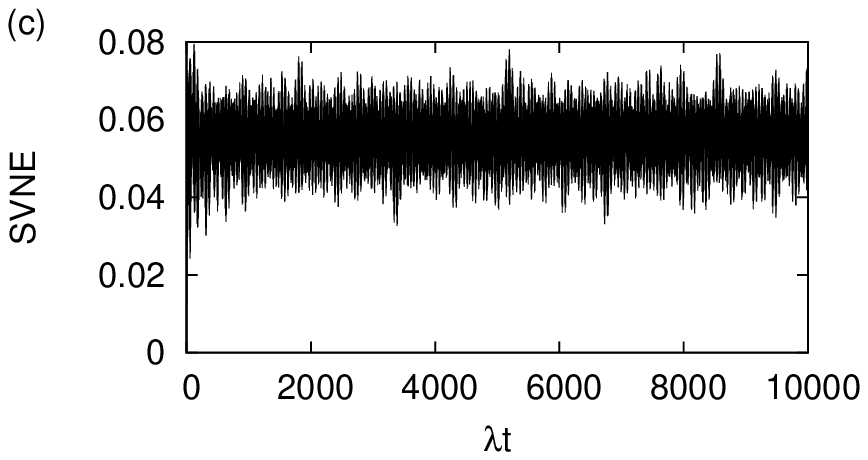}
\caption{$S_{\rm a}(t)$ in the presence 
of squeezing.  Initial states  
(a) $\ket{1;\xi;\alpha,5}$, (b) $\ket{1;\xi; \alpha, 10}$ and (c) $\ket{1;\xi;\xi}$, 
 with ${|\alpha|}^{2} =10$ and 
 squeezing parameter $\xi = 2$. In all cases, $\Delta = 0$ and  $\chi/\lambda = 5$.} 
\label{fig:svne_4}
\end{figure}

For completeness, we have also investigated the role of squeezing on entanglement collapse in this model. No collapse is exhibited by $S_{\rm a}$  when $F_{1}$ 
is a standard squeezed vacuum state (labelled by the complex parameter $\xi$) and $F_{2}$ is either a PACS or a squeezed vacuum 
state (with the same parameter $\xi$). Figures \ref{fig:svne_4}(a) and (b) 
illustrate these conclusions. Moreover, we have 
verified that, for those initial  states for which 
a collapse of $S_{\rm a}$ does occur, the field states  do not exhibit squeezing or second-order squeezing during the time interval of the collapse. 
It would seem, therefore, that it is the extent of {\em coherence}, above all else, 
 that is the primary 
determining factor in the occurrence of entanglement collapses in our 
model system. 

\subsection{Intensity-dependent couplings: From SVNE collapse to the revival phenomenon}
\paragraph{} 

We turn,now, to the  effect of   
 an intensity-dependent  coupling of the atom with the field 
 modes, as characterised by 
 a non-constant function $f(N_{i})$ in (\ref{defnofR}). 
 
  \begin{figure}[h]
\centering
\includegraphics[scale=0.57]{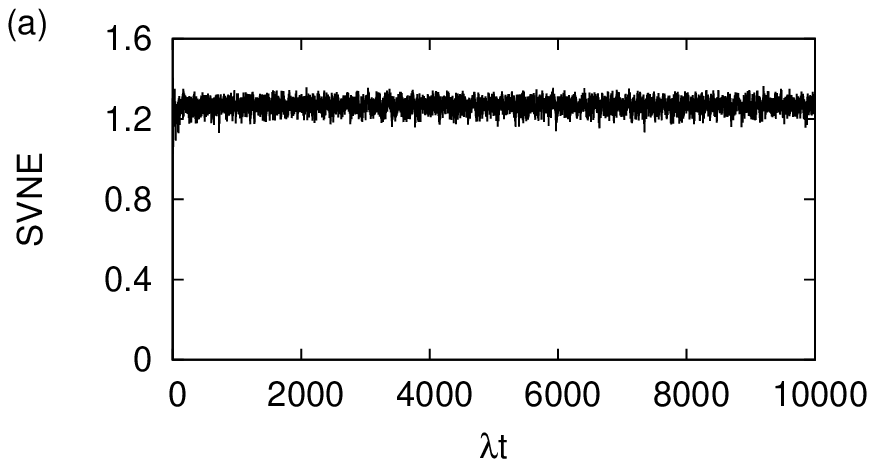}
\includegraphics[scale=0.57]{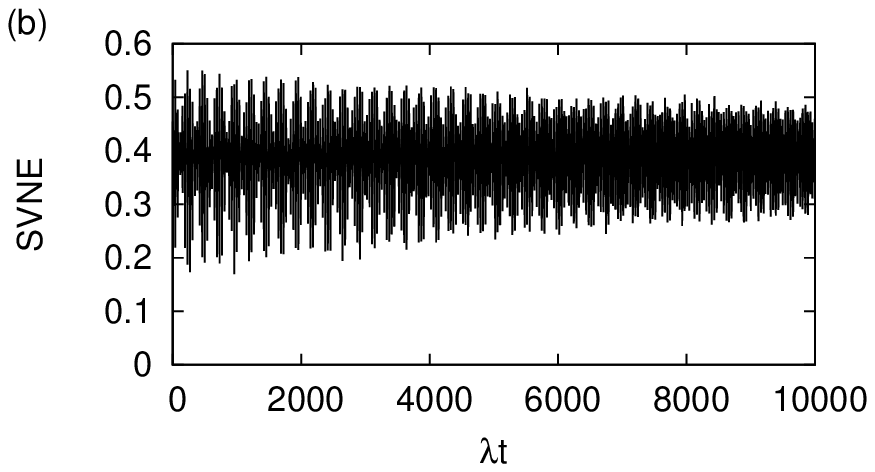}
\includegraphics[scale=0.57]{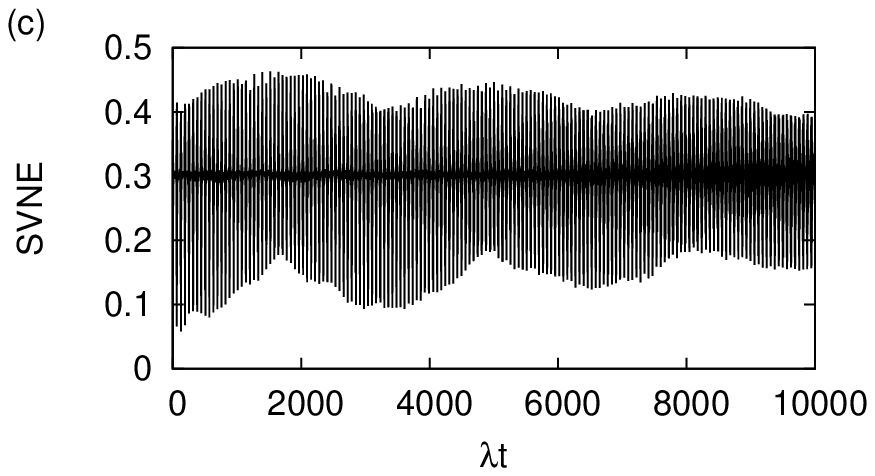}
\caption{$S_{\rm a}(t)$  for 
 intensity-dependent coupling $f(N) = N^{1/2}$.  Initial  state 
$\ket{1;\alpha, 5; \alpha, 5}$, 
${|\alpha|}^{2} = 10$, $\Delta = 0$. (a) $\chi = 0$, 
(b) $\chi/\lambda = 6$,  (c) $\chi/\lambda = 10$.} 
\label{fig:svne_6}
\end{figure}
\begin{figure}[h]
\centering
\includegraphics[scale=0.57]{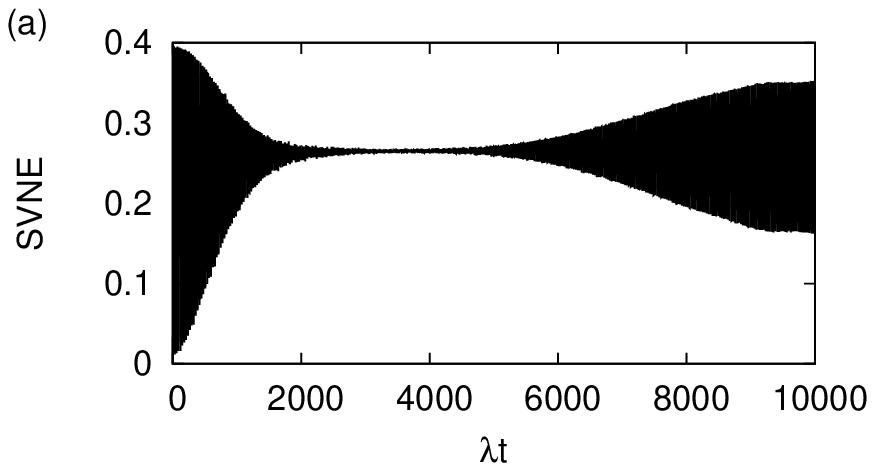}
\includegraphics[scale=0.57]{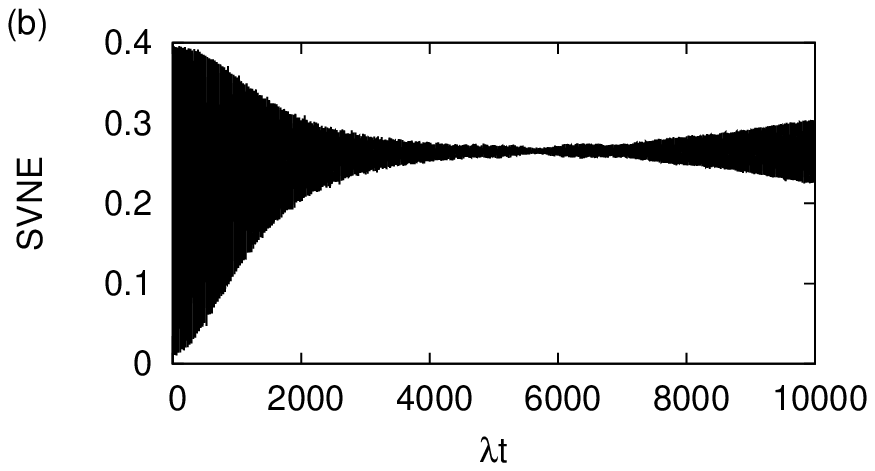}
\includegraphics[scale=0.57]{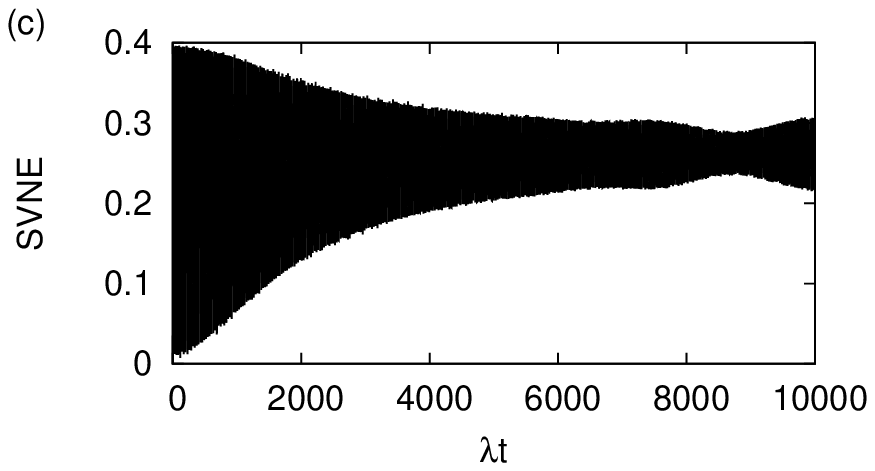}
\includegraphics[scale=0.57]{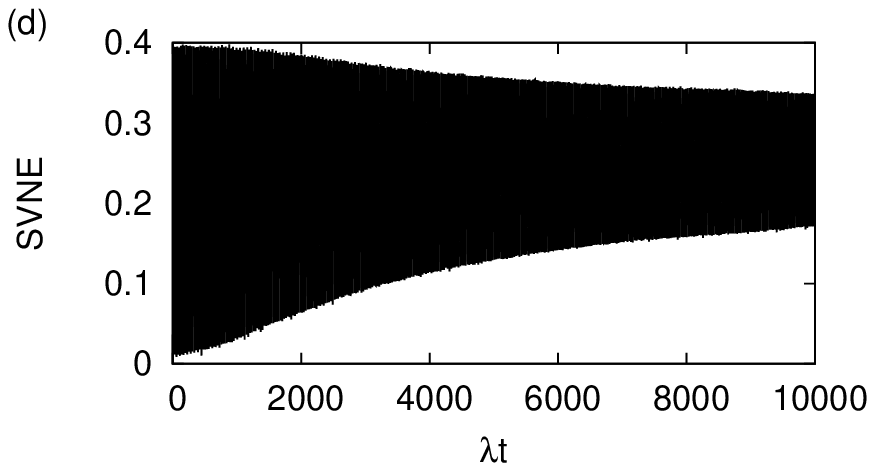}
\includegraphics[scale=0.57]{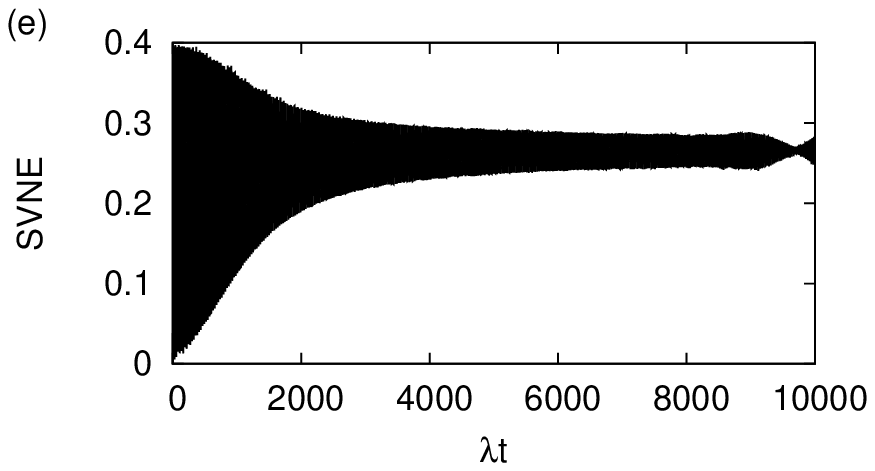}
\includegraphics[scale=0.57]{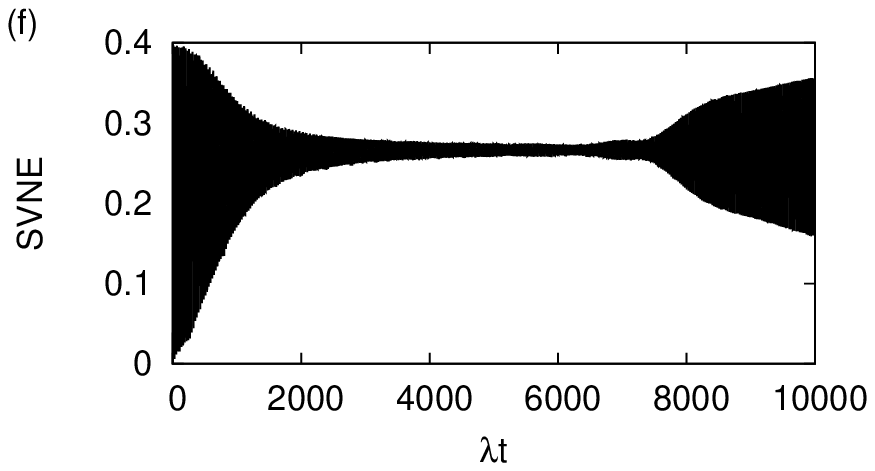}
\includegraphics[scale=0.57]{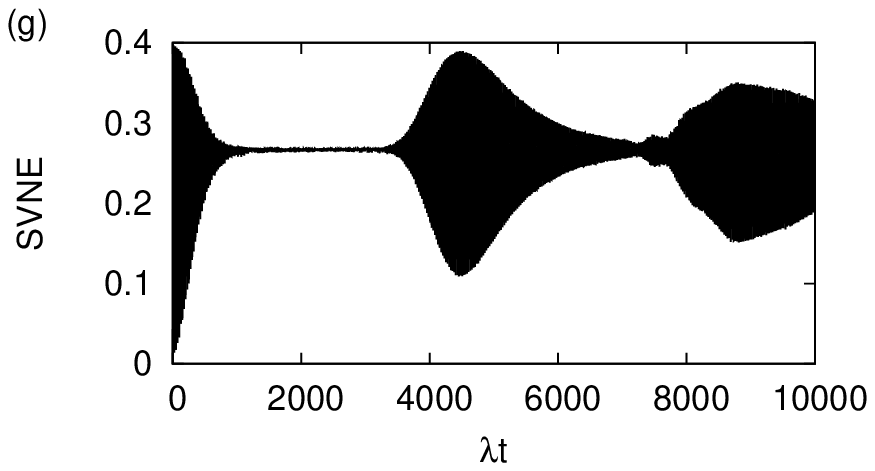}
\includegraphics[scale=0.57]{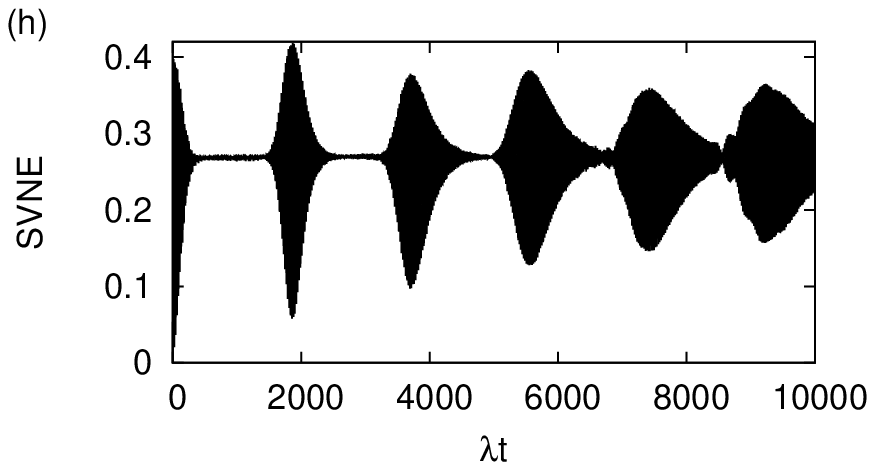}
\includegraphics[scale=0.57]{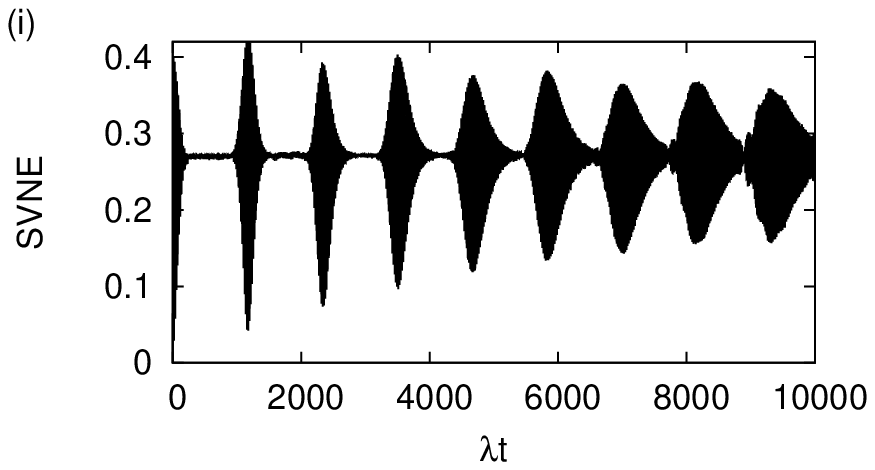}
\includegraphics[scale=0.57]{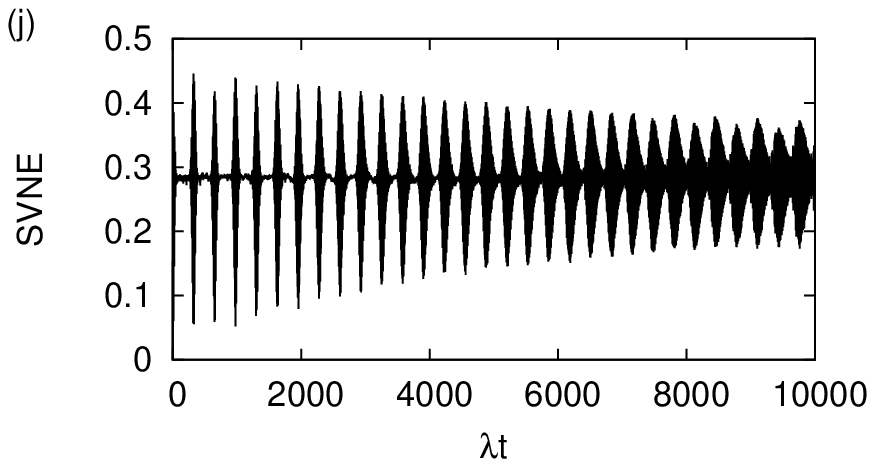}
\includegraphics[scale=0.57]{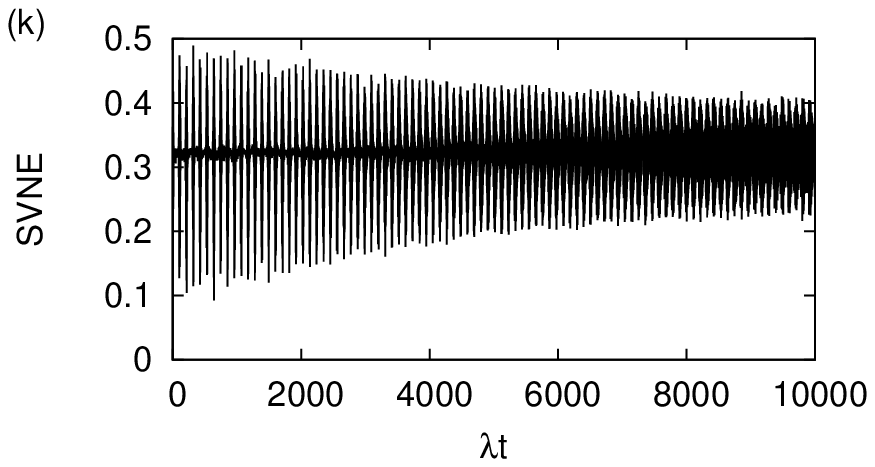}
\includegraphics[scale=0.57]{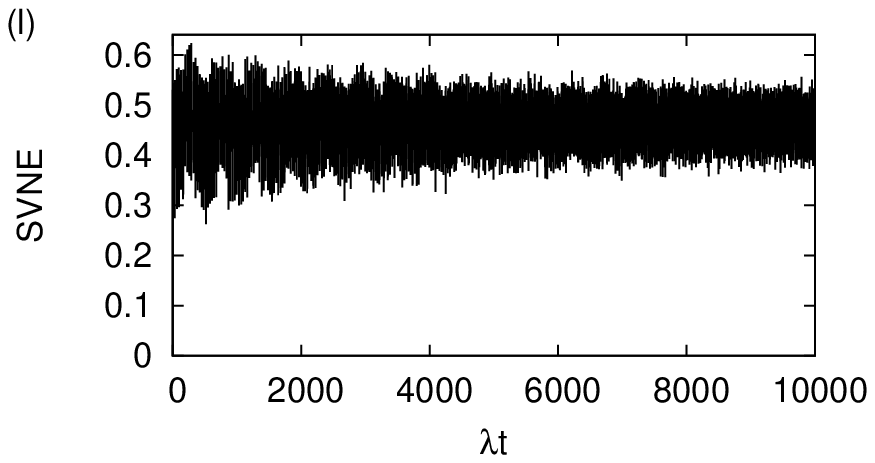}
\caption{$S_{\rm a}(t)$ for
 intensity-dependent coupling $f(N) = (1+\kappa N)^{1/2}$. 
Initial state $\ket{1;\alpha, 5; \alpha, 5}, \, |\alpha|^{2} = 10, \,\Delta = 0, \,\chi/\lambda = 5$. 
The value of $\kappa$  is (a) 0, (b)  0.0012, (c) 0.002, (d) 0.0034, (e) 0.005, (f)  0.006, (g) 0.01, (h) 0.02, (i) 0.03, (j) 0.1, (k) 0.3 and (l) 1.} 
\label{fig:svne_5}
\end{figure}

In order  to  facilitate ready  comparison,  we consider the same initial state, 
$\ket{1;\alpha, 5; \alpha, 5}$ with $|\alpha|^{2} = 10$ and  $\Delta = 0$, 
  as in figures \ref{fig:svne_1}(a),  
    \ref{fig:svne_3}(a) and  \ref{fig:svne_3}(b). These   
 correspond to intensity-independent coupling, i.e.,  $f(N_{i}) = 1$, for 
 a range of values of $\chi/\lambda$.  Consider,  now, the functional form\cite{buck} 
 $f(N_{i}) = N_{i}^{1/2}$.
 Figures \ref{fig:svne_6}(a)-(c) depict the behaviour of 
 $S_{\rm a}(t)$  in this case. 
 It is evident that SVNE collapse is absent now,  even for strong nonlinearity. 
  The behaviour in these two limiting cases motivates  an examination 
of the effect of the intensity-dependent coupling 
$f(N_{i}) = (1+\kappa N_{i})^{1/2}$ 
 over a range of values of the 
 parameter $\kappa$, running from $0$ to 
 $1$.  Remarkably diverse  features  emerge, including SVNE collapse 
 as well as  recurrent collapses and revivals,  as shown in 
 figures \ref{fig:svne_5}(a)-(l). The behaviour is sensitively dependent on the 
 value of $\kappa$. 

The collapse of the SVNE over a considerable time interval 
 that occurs  for $\kappa = 0$  (figure \ref{fig:svne_5}(a)) is 
 gradually lost (figures 
\ref{fig:svne_5}(b) to (e)) as $\kappa$ increases 
to a value slightly  above $0. 005$ . An apparent precursor to  collapse  
again appears as $\kappa$ is further increased slightly 
(figure  \ref{fig:svne_5}(f), $\kappa = 0.006$), but  what  happens is that there is a 
shorter-duration collapse, followed by a revival, and then a 
second incipient collapse and revival (figure \ref{fig:svne_5}(g), $\kappa = 0.01$), 
within the original time interval of collapse. A further small increase in $\kappa$ 
produces a distinctive  sequence of collapses and revivals 
(figures \ref{fig:svne_5}(h), (i), (j)), which remains clear-cut till 
$\kappa$ becomes a little larger than  $0.1$. Subsequently, the 
intervals between successive revivals become too short to be 
discernible on the scale used in figure \ref{fig:svne_5}, and, moreover, 
fractional revivals start filling these small intervals (figure \ref{fig:svne_5}(k), 
$\kappa = 0.3$). As $\kappa$ is increased further, the entanglement 
collapse and revival phenomenon 
is no longer discernible. Figure \ref{fig:svne_5}(l) depicts $S_{\rm a}(t)$ 
for $\kappa = 1$. It is thus evident that a series of qualitative changes 
is exhibited by the entanglement entropy in the system under 
study as the value of $\kappa$ is changed  in a relatively small 
range, signalling very sensitive dependence on this parameter 
in a manner reminiscent of bifurcation cascades preceding the 
onset of chaos  in nonlinear classical 
dynamical systems.

Finally, as stated in the Introduction, we have verified that these 
interesting features are also exhibited in the case of a V atom interacting with two radiation modes.

\section{Concluding remarks}
\label{concluding remarks}
\paragraph{} 
We have considered  a tripartite system comprising  
a $\Lambda$-type or V-type atom interacting with two radiation fields. The mean photon number corresponding to the probe field is seen to display collapses and revivals  for specific initial field states and parameters in the Hamiltonian, in the absence of the coupling field.  With the coupling field  turned on, a window of electromagnetically-induced 
transparency appears during the collapse interval in the absorption spectrum. 
On a longer time scale,  interesting dynamical effects are observed 
in  the time evolution of the entanglement.  This includes  a collapse of the
subsystem von Neumann entropy 
of the atom over a considerable  time interval.  Both these features 
are  sensitive to the nature of the initial states of the fields,  and seem to 
reflect the extent of the departure from coherence of those states.
 
We have  attempted to identify the roles played by field nonlinearities, detuning parameters and the departure from coherence of the initial states of the radiation fields on EIT and on the extent of entanglement between subsystems during temporal evolution,  since 
these two aspects are  suitable  for potential experimental investigations.  Detailed experiments on EIT, the identification of the photon-added coherent state in the laboratory,  and  the necessity of retaining  the extent of entanglement between subsystems from the point of view of quantum information processing  add impetus to our investigations.  Reconstruction of the state of the system during SVNE collapse could possibly be attempted  through continuous-variable quantum state tomography. The unanticipated behaviour of  the SVNE  as the strength of the intensity-dependent field-atom  coupling  is varied would correspondingly take the system  through a spectrum of nonclassical states which  are worth identifying through state reconstruction procedures. The mean photon number is seen to reflect the long time dynamics of the SVNE corresponding to the atomic subsystem. This is an observable which lends itself to experimental 
observation,  and hence one that can  be used to examine the 
dynamical features predicted for the tripartite model considered.

\end{document}